\documentclass[twocolumn,showpacs,amsmath,amssymb,prl]{revtex4}
\usepackage{graphicx}

\begin{document}

\title{  Tuning electronic properties of FeSe$_{0.5}$Te$_{0.5}$ thin flakes by a novel field effect transistor}
\author{C. S. Zhu$^{1,}$\footnotemark[2], J. H. Cui$^{1,}$\footnotemark[2], B. Lei$^1$, N. Z. Wang$^1$, C. Shang$^1$, F. B. Meng$^1$, L. K. Ma$^1$, X. G. Luo$^{1,4}$, T. Wu$^{1,4}$, Z.
Sun$^{2,4}$,}
\author{X. H. Chen$^{1,3,4}$}
\email{chenxh@ustc.edu.cn}

\affiliation{ $^1$Hefei National Laboratory for Physical Sciences at
Microscale and Department of Physics, and CAS Key Laboratory
of Strongly-coupled Quantum Matter Physics, University of Science and
Technology of China, Hefei, Anhui 230026, China\\
$^2$National Synchrotron Radiation Laboratory, University of Science
and Technology of China, Hefei, Anhui 230026, China\\
$^3$High Magnetic Field Laboratory, Chinese Academy of Sciences,
Hefei, Anhui 230031, China\\
$^4$Collaborative Innovation Center of Advanced Microstructures,
Nanjing University, Nanjing 210093, China}

\date{\today}

\begin{abstract}
Using a field-effect transistor (FET) configuration with solid
Li-ion conductor (SIC) as gate dielectric, we have successfully tuned
carrier density in FeSe$_{0.5}$Te$_{0.5}$ thin flakes, and the
electronic phase diagram has been mapped out. It is found that electron
doping controlled by SIC-FET leads to a suppression of the
superconducting phase, and eventually gives rise to an insulating
state in FeSe$_{0.5}$Te$_{0.5}$. During the gating process, the
(001) peak in XRD patterns stays at the same position and no new
diffraction peak emerges, indicating no evident Li$^+$ ions
intercalation into the FeSe$_{0.5}$Te$_{0.5}$. It indicates that a
systematic change of electronic properties in FeSe$_{0.5}$Te$_{0.5}$
arises from the electrostatic doping induced by the accumulation of
Li$^+$ ions at the interface between FeSe$_{0.5}$Te$_{0.5}$ and
solid ion conductor in the devices. It is striking that these
findings are drastically different from the observation in FeSe thin
flakes using the same SIC-FET, in which $T_c$ is enhanced from 8 K
to larger than 40 K, then the system goes into an insulating phase
accompanied by structural transitions.
\end{abstract}



\maketitle

\footnotetext[2]{These authors contributed equally to this work.}

Tuning carrier concentration is one of the most powerful approaches
in the condensed matter physics for the explorations of novel
quantum phases and exotic electronic properties as well as their
underlying physical mechanics \cite{1,2,3,4,5,6,7,8}. To overcome
the inherent doping limit in the material synthetic methods, field
effect transistor (FET) configurations have been applied to tune
material properties using gating by electric field \cite{9}. Two
types of FET, metal-insulator-semiconductor (MIS) FET and electric
double layer (EDL) FET, are widely used to
control the charge carrier density on the surface of
materials \cite{10,11}. In order to change the carrier density in the
bulk, the so-called ionic field-effect transistor (iFET) with
gel-like electrolyte as the gate medium has been used to drive
Li$^+$ ions into layered materials. This type of FET configuration
can effectively modulate 1T-TaS$_{2}$ electronic properties by the
tunable Li$^+$ ion intercalation \cite{12}. However, the heavily-doped
electronic states in all these FET configurations are confined at
the interfaces or overlaid with electrolyte, which prevents them
from being characterized by many physical measurements. On the other
hand, conventional MIS-FET devices cannot provide sufficient
carriers to induce novel phases, such as superconductivity, by
electrostatic doping, and the liquid or gel-like electrolyte is not
compatible with modern solid electronics and may react with samples
when gating voltage is applied \cite{11,13,14}. Recently, we
have fabricated a new type of FET device (see Fig. 1(c)) using solid ion
conductor (SIC) as a gate dielectric. This type of FET configuration
overcomes the inherent limitations mentioned above, and its
application on FeSe thin flakes reveals some substantial advantages
of the SIC-FET \cite{15}.

Using the SIC-FET devices, we are able to tune the carrier density
of FeSe by driving lithium ions in and out of the FeSe thin flakes
and thus control the material properties and its phase transitions.
With the intercalation of Li$^+$ ions, new structural changes have
been identified by \textit{in-situ} XRD measurements. A wide carrier-doping
phase diagram has also been mapped out with increasing Li$^+$ content. A
dome-shaped superconductivity with maximal $T_c$ above 40 K was
obtained and an insulating phase was reached at extremely overdoped
regime. In addition, we have demonstrated that many experimental
probes can be applied to uncover novel structural phases that are
inaccessible in ordinary conditions \cite{15}. The
application of such a novel FET device can provide exciting
opportunities for exploring new quantum phases and novel materials.

\begin{figure}[h]
    \centering
    \includegraphics[width=0.45\textwidth]{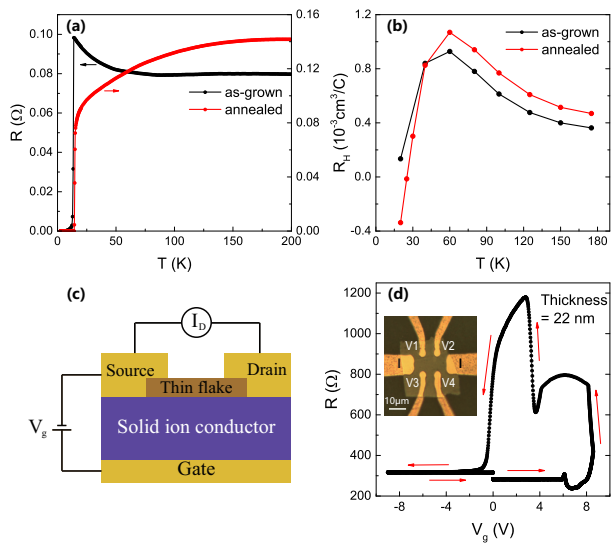}
    \caption{(color online). (a): Temperature-dependent resistivity of as-grown and O$_2$-annealed FeSe$_{0.5}$Te$_{0.5}$ bulk single crystals.
    (b): Temperature-dependent Hall coefficient $R_H$ of as-grown and O$_2$-annealed FeSe$_{0.5}$Te$_{0.5}$ thin flakes. (c): A schematic plot of the
    SIC-FET device with the solid ion conductor as the gate dielectric. (d): A gate-voltage dependence of the resistance of FeSe$_{0.5}$Te$_{0.5}$ thin
    fakes with thickness of 22 nm. The insert is the optical image of a FeSe$_{0.5}$Te$_{0.5}$ thin flake with the current and voltage terminals labeled. }
    \label{Fig. 1}
\end{figure}

FeTe and FeSe share the same crystal structure. The former has very
strong electron correlation, while the latter shows only moderate
electron correlation \cite{16}. The isovalent substitution of Te for
Se results in a remarkable change of electron correlation
\cite{16,17}. Transport measurements have shown that, upon
substitution, $T_c$ increases from 8 K in FeSe to the maximum of
$\sim$ 14 K around $x$ $\sim$ 0.5 in FeSe$_{1-x}$Te$_{x}$, and bulk
superconductivity disappears for $x$ $>$ 0.7 \cite{18,19,20,21,22,23}.
In FeSe and its derived compounds, $T_c$ can be significantly
increased up to more than 40 K when electron carriers are introduced
into the bulk \cite{24,25}. On the basis of our previous studies on
FeSe thin flakes, it is interesting to investigate whether the $T_c$
of FeSe$_{1-x}$Te$_{x}$ can also be significantly improved by
introduction of electronic carrier using the same SIC-FET approach
and how the enhanced electron correlation with Te doping influences
the superconductivity.

In this paper, we use the SIC-FET device to investigate how electron
doping can modify the electronic properties of
FeSe$_{0.5}$Te$_{0.5}$. In contrast to the case of FeSe \cite{15},
we found that Li$^+$ intercalation is not evident in
FeSe$_{0.5}$Te$_{0.5}$ and that there is no structural transition
during the gating process. However, the electronic properties can
still be modulated, suggesting that Li$^+$ ions accumulate at the
interface between the thin flake and the SIC substrate and induce
electrostatic doping in FeSe$_{0.5}$Te$_{0.5}$. Though electron-type
carriers are doped into FeSe$_{0.5}$Te$_{0.5}$ system, Hall
measurements indicate that hole-type carriers always dominate the
transport properties at high temperatures. With increasing the gate
voltage, the superconducting state is suppressed, and eventually an
insulating phase emerges, which could be related to the strong
electron correlation in FeSe$_{0.5}$Te$_{0.5}$. These findings are
strikingly different from the observation of FeSe thin flakes using
the same SIC-FET device, in which the $T_c$ can be enhanced from 8 K
(low $T_c$ phase) to more than 40 K (high $T_c$ phase).

\begin{figure}[h]
    \centering
    \includegraphics[width=0.45\textwidth]{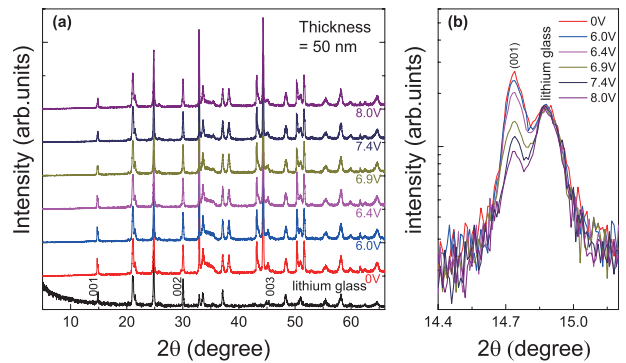}
    \caption{(color online). (a) The typical \textit{in-situ} XRD patterns of the FeSe$_{0.5}$Te$_{0.5}$ thin fakes with thickness of 50 nm at various gating
    status. (b) The magnified view of the (001) diffraction peak. }
    \label{Fig. 2}
\end{figure}

The as-grown FeSe$_{0.5}$Te$_{0.5}$ crystals are superconducting and
reach zero-resistance at T = 11 K (Fig. 1(a)). The non-metallic
characteristic is caused by the localization effect from
interstitial Fe \cite{26,27}. After removing the interstitial Fe by
annealing crystals in oxygen, we improved the quality and enhanced
the $T_c^{zero}$ to 14.2 K. In Fig. 1(b), $R_H$ shows a sign
reversal below 30K in the annealed crystals, suggesting that
electron-type carriers become intrinsically dominant at low
temperatures. Using the annealed crystals and following the details
as described in ref. \cite{15}, we fabricated SIC-FET devices, and
performed resistance, Hall coefficient, and \textit{in-situ} XRD
measurements.

Figure 1(c) shows a schematic plot of the SIC-FET device we used to
tune the electronic properties of FeSe$_{0.5}$Te$_{0.5}$ thin
flakes. Using the solid state lithium ion conductive glass ceramics
as the gate dielectric, we prepared devices with a standard Hall bar
configuration (inset of Fig. 1(d)). Thin flakes with a typical
thickness of 22 nm serve as the transport channel. Li$^+$ in
the lithium ion conductor can be driven by electric field. With a
positive gate voltage applied, Li$^+$ accumulates at the bottom of
FeSe$_{0.5}$Te$_{0.5}$ thin flakes. As will be shown later, our XRD
data and transport measurements suggest that Li$^+$ intercalation is
minimal with further increasing the gate voltage. However, the
electronic properties of FeSe$_{0.5}$Te$_{0.5}$ can still be
drastically modulated by gating.

Figure 1(d) shows a typical $R$-$V_g$ curve taken at T = 260 K. The
resistance of the FeSe$_{0.5}$Te$_{0.5}$ thin flake stays at the
same value when $V_g$ $<$ 5 V. With increasing $V_g$ above 5 V, the
resistance varies and the details will be shown in Fig. 3. As the gate voltage is sweeping back, the
resistance increases drastically by a factor of 4 and drops
drastically when the gate voltage is descending to 0 V. Eventually it returns to a value close to its initial status with
negative voltage applied. This behavior indicates that the modulation of electronic properties by gating is
reversible. The slight different resistance of
FeSe$_{0.5}$Te$_{0.5}$ before and after gating may arise from the
degradation of the bottom surface of thin flakes.

\begin{figure}[h]
    \centering
    \includegraphics[width=0.45\textwidth]{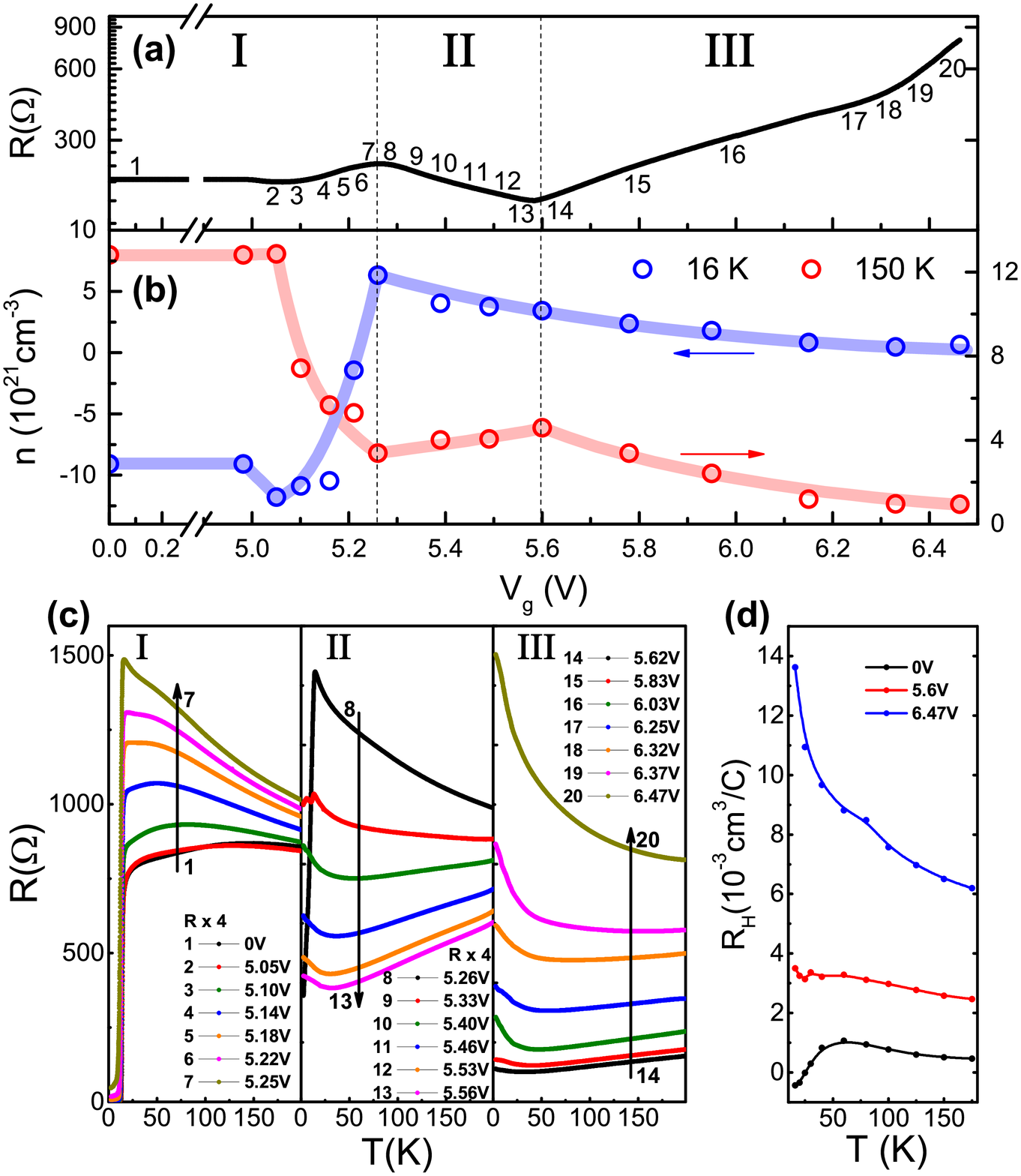}
    \caption{(color online). (a): Gate voltage dependence of the resistance of a FeSe$_{0.5}$Te$_{0.5}$
 thin flake with a thickness of 22 nm. (b):  Hall number $n_H$ with
various gate voltage applied, taken at T = 16 K and 150 K. The low
temperature $n_H$ shows a sudden sign reversal around $V_g$ = 5.22
V. (c): The longitudinal resistance at various gating status. (d):
Temperature dependence of Hall coefficient $R_H$, calculated from
the linear fit of a $R_{xy}$(B) plot from -9 to 9 T. For the
nonlinear $R_{xy}$(B) curves, $R_H$ was extracted from the slope of
the high-field quasilinear part \cite{15}.} \label{Fig. 3}
\end{figure}

By performing \textit{in-situ} XRD measurements on
FeSe$_{0.5}$Te$_{0.5}$ devices, we studied whether the structure
changes during the gating process. The 22 nm thin flakes are too
thin to yield detectable signals, and we thus selected 50 nm
FeSe$_{0.5}$Te$_{0.5}$ thin flakes for XRD characterization. Fig.
2(a) shows the typical XRD patterns at different charge stages. The
(001) diffraction peak stays at the same position with increasing
gate voltage (Fig. 2(b)), indicating that the interlayer separation is
not affected by gating and that Li$^+$ intercalation should be
minimal if it does occur. This behavior is different from FeSe
thin flakes, in which Li$^+$ intercalation is substantial and causes
structural transitions. The decrease of the (001) peak intensity
could be attributed to the degradation of FeSe$_{0.5}$Te$_{0.5}$ bottom
layer due to Li$^+$ accumulation, which can weaken the diffraction
signals.

Though Li$^+$ intercalation is restrained, the accumulation of
Li$^+$ ions at the interface can also cause electrostatic doping in
FeSe$_{0.5}$Te$_{0.5}$ thin flakes. Fig. 3(a) shows the details of
the evolution of resistance with gate voltage. The variation of
resistance changes slightly for different devices, while the
dip-peak-valley feature and the rise of resistance at higher gate
voltages are highly reproducible. To further demonstrate carrier
doping, we performed Hall measurements. In Fig. 3(b), the Hall
number $n_H$ = 1/e$R_H$ measured at T = 150 K and T = 16 K varies
drastically, suggesting that the carrier concentration in the bulk
is strongly affected during gating. Based on the transport
measurements, we empirically divide the charge doping process into
three regions I, II and III, and plot the temperature-dependent
resistance in Fig. 3(c). In the I region, the thin flakes show
superconducting behavior, and the $T_c$ is slightly suppressed with
increasing the gate voltage. In this region, the corresponding
low-temperature Hall number is negative, indicating that electron
carriers  are crucial for superconductivity. In the II region, the
superconductivity is quickly suppressed and a semiconducting-like
feature appears at low temperature. Above 30 K, from I to III
region, the overall normal-state resistance increases, decreases and
increases again, accompanied by several crossovers between metallic
and semiconducting-like behavior.

Figure 3(d) plots the temperature dependence of the Hall coefficient
$R_H$ with different gate voltages. Except for the sign reversal behavior at
low temperature in the I region, the transport properties of
FeSe$_{0.5}$Te$_{0.5}$ are dominated by hole-type carriers. The overall $R_H$ value
rises strongly with increasing the gate voltage, indicating that electron doping
in the thin flake is induced by gating to counteract the contribution from the hole-type
carriers, though Fig. 3(b) shows that the behavior in the II region appears slightly anomaly at T=150 K.
Generally speaking, as shown in our studies on FeSe thin flakes, Li$^+$ intercalation can
bring a great amount of electrons into the system, however the Hall number in the II and III
regions indicates that the hole carriers are always dominant in the gating range we studied
here. These results serve as a clear evidence that Li$^+$ intercalation is minimal in our devices and cannot dope enough electrons.

\begin{figure}[h]
    \centering
    \includegraphics[width=0.45\textwidth]{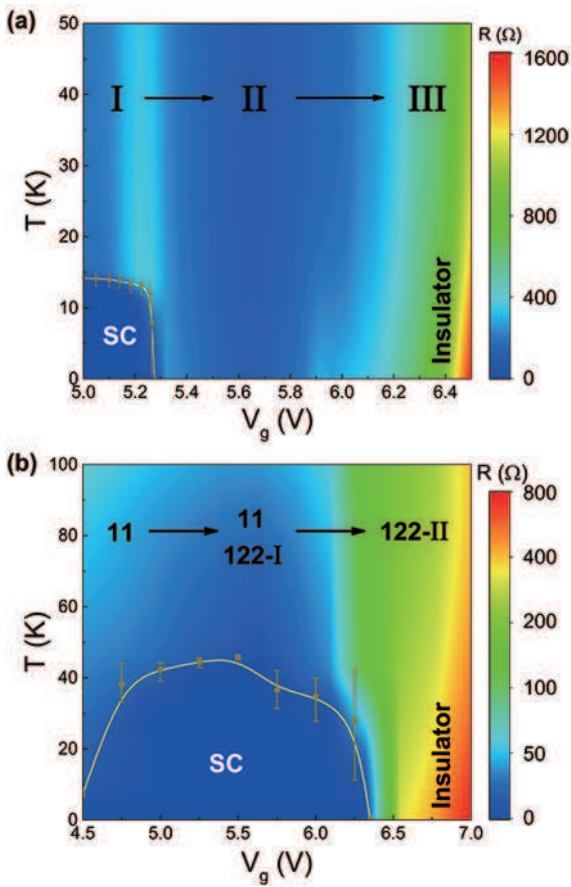}
    \caption{(color online). (a): The phase diagram of FeSe$_{0.5}$Te$_{0.5}$ thin flake as a function of gate voltage. (b): The phase diagram of
        Li-intercalated FeSe thin fake as a function of gate voltage, and the data was taken from Ref. 15.}
    \label{Fig. 4}
\end{figure}

Based on the resistance measurements, we plot an electronic phase
diagram as a function of gate voltage for FeSe$_{0.5}$Te$_{0.5}$
(see Fig. 4(a)). The electronic phase diagram of FeSe is also shown
for comparison (Fig. 4(b)). In FeSe$_{0.5}$Te$_{0.5}$, as
$V_g$ increases, a series of variations from superconducting phase to insulating phase take place. The evolution
from superconductor to insulator induced by electron doping shares
similar behavior with FeSe thin flakes. Nevertheless, one may notice
two remarkable differences between FeSe$_{0.5}$Te$_{0.5}$ and FeSe.
First, Li$^+$ ions can be easily intercalated into FeSe thin flakes,
and consequently structural phase transitions have been observed \cite{15}.
However, Li$^+$ intercalation is minimal in FeSe$_{0.5}$Te$_{0.5}$
and there is no evident change in structure. This difference
suggests that the substitution of Te for Se significantly changes
the characteristics of two-dimensional FeSe layers and restrains the
intercalation of Li$^+$ ions. Secondly, electron doping can
drastically increase the $T_c$ from $\sim$8 K to more than 40 K for
FeSe \cite{15}, but suppresses the $T_c$ from $\sim$ 14 K down to lower
value instead of enhancing it for FeSe$_{0.5}$Te$_{0.5}$. This
contrast suggests that the substitution of Te for Se severely
changes the electronic properties, though the crystal structures of
these two compounds are highly similar to each other.

Direct probing electronic structures by angle-resolved photoemission
spectroscopy (ARPES) has shown that FeSe possesses moderate electron
correlation with well-defined electronic bands, while  the bands in
FeSe$_{1-x}$Te$_{x}$ are heavily renormalized for x $\geq$ 0.5,
indicating strong electron correlation in this substitution region
\cite{16,17}. This difference is likely responsible for the
discrepancy of carrier doping-induced electronic properties in these
two materials. The pristine FeSe has both of electron and hole
pockets at the Fermi level with a relatively low $T_c$ of 8 K. With
heavy electron doping, the Fermi level shifts up drastically and
electron pockets expand, the resultant electronic system shows a
very high  $T_c$ more than 40 K. ARPES measurements of FeSe$_{0.5}$Te$_{0.5}$ show unambiguous
hole pockets around the zone center ${\Gamma}$ in \textit{k}-space,
while the electron pocket around the zone corner M is ill-defined. This characteristic explains why the
hole-type carriers dominate significantly in Hall measurements at
high temperatures.  A detailed analysis of the ill-defined electron
pockets suggests that the coherent spectral weight gradually
increases as temperature is decreasing \cite{28}, which is naturally
related to the sign reversal of Hall $n_H$ at low temperatures. In
FeSe$_{0.5}$Te$_{0.5}$, our data suggests that Li$^+$ intercalation
is not evident and thus the electron doping is not so strong as the
case in FeSe. Therefore, the consequent electronic structures after
gating are remarkably different in FeSe and  FeSe$_{0.5}$Te$_{0.5}$.
This characteristic is consistent with the fact that hole-type
carriers still contribute significantly to transport properties as
high gate voltage is applied.

In Fig. 3(b), the high-temperature Hall data in the I region
indicates that the electron doping can effectively counteract the
hole-type carriers in transport properties. However, the evolution
of Hall $n_H$ at 16 K suggests that the contribution from electron-type carriers is
also suppressed by electron doping. This anomalous trend is inconsistent with a rigid band model, in which electron doping is
expected to increase electron-type carriers.
We argue that such an anomaly should be
associated with some novel change of electron pockets around M, in
which a gap or a strong suppression of spectral weight may occur due
to the increase of electron doping and results in a reduction of electron-type contribution in Hall $n_H$.
With increasing the gate voltage, eventually both of electron-type and hole type carriers are suppressed
and the whole system shows an insulating behavior. If we reverse the carrier doping procedure from the insulating
phase, the evolution of the electronics states will show that a practical hole doping
leads to superconductivity. Such a
behavior shares similarity with cuprates, in which
the superconductivity is realized by doping holes into a Mott
insulator. From this perspective, we argue that there is novel
physics of strong electron correlation during the change of electronic properties
induced by gating. Further investigations of the transitions between
different phases will enrich our understanding of the relationship
between the superconductivity and the correlated electron state
in the insulating region.

In summary, we have successfully modulated electronic properties of
FeSe$_{0.5}$Te$_{0.5}$ thin flakes using SIC-FET configuration.
Contrary to our previous studies of FeSe thin flakes, XRD patterns
indicate that Li$^+$ intercalation is not evident in our
FeSe$_{0.5}$Te$_{0.5}$ FET devices. Our data suggests that, Li$^+$
ions likely accumulate with the gate voltage applied at the
interface between FeSe$_{0.5}$Te$_{0.5}$ and solid ion conductor,
which can lead to electrostatic doping and is responsible for the
drastic change of electronic properties of FeSe$_{0.5}$Te$_{0.5}$. A
systematic variation of transport properties in
FeSe$_{0.5}$Te$_{0.5}$ reveals a change of electronic states from a
superconducting state to an insulating state, which can be related
to the strong electron correlation in this material. The
substitution of Te for Se significantly changes the characteristics
of two-dimensional FeSe layers, and enhances the electronic correlation.\\

This work is supported by the National Natural Science Foundation of
China (Grants No. 11190021, No. 11534010 and No.
91422303), the Strategic Priority Research Program (B) of the
Chinese Academy of Sciences (Grant No. XDB04040100), the National Key R\&D
Program of the MOST of China (Grant No. 2016YFA0300201), and the Hefei
Science Center CAS (2016HSC-IU001).


\begin{thebibliography}{99}


\bibitem{1}A. D. Caviglia, S. Gariglio, N. Reyren, D. Jaccard, T. Schneider, M. Gabay, S. Thiel, G. Hammerl, J. Mannhart, and J.-M. Triscone, Nature
\textbf{456}, 624-627 (2008).
\bibitem{2}K. Ueno, S. Nakamura, H. Shimotani, A. Ohtomo, N. Kimura, T. Nojima, H. Aoki, Y. Iwasa, and M. Kawasaki, Nat. Mater.
\textbf{7}, 855-858 (2008).
\bibitem{3}K. Ueno, S. Nakamura, H. Shimotani, H. T. Yuan, N. Kimura, T. Nojima, H. Aoki, Y. Iwasa, and M. Kawasaki, Nat. Nanotechnol.
\textbf{6}, 408-412 (2011).
\bibitem{4}J. T. Ye, S. Inoue, K. Kobayashi, Y. Kasahara, H. T. Yuan, H. Shimotani, and Y. Iwasa, Nat. Mater.
\textbf{9}, 125-128 (2010).
\bibitem{5}A. T. Bollinger, G. Dubuis, J. Yoon, D. Pavuna, J. Misewich, and I. Bozovic, Nature
\textbf{472}, 458-460 (2011).
\bibitem{6}Y. Saito, Y. Kasahara, J. T. Ye, Y. Iwasa, and T. Nojima, Science
\textbf{350}, 409-413 (2015).
\bibitem{7}L. J. Li, E. C. T. O'Farrell, K. P. Loh, G. Eda, B. Ozyilmaz, and A. H. Castro Neto, Nature
\textbf{529}, 185-189 (2016).
\bibitem{8} B. Lei, Z. J. Xiang, X. F. Lu, N. Z. Wang, J. R. Chang, C. Shang, A. M. Zhang, Q. M. Zhang, X. G. Luo, T. Wu, Z. Sun, and X. H. Chen, Phys. Rev. B
\textbf{93}, 060501(R) (2016).
\bibitem{9}C. H. Ahn, S. Gariglio, P. Paruch, T. Tybell, L. Antognazza, and J.-M. Triscone, Science
\textbf{284}, 1152-1155 (1999).
\bibitem{10}C. H. Ahn, A. Bhattacharya, M. Di Ventra, J. N. Eckstein, C. Daniel Frisbie, M. E. Gershenson, A. M. Goldman, I. H. Inoue, J. Mannhart, and J. Triscone, Rev. Mod. Phys.
\textbf{78}, 1185-1212 (2006).
\bibitem{11}K. Ueno, H. Shimotani, H. T. Yuan, J. T. Ye, M. Kawasaki, and Y. Iwasa, J. Phys. Soc. Jap.
\textbf{83}, 032001 (2014).
\bibitem{12}Y. J. Yu, F. Y. Yang, X. F. Lu, Y. J. Yan, Y. H. Cho, L. G. Ma, X. H. Niu, S. Kim, X. H. Chen, and Y. B. Zhang, Nat. Nanotechnol.
\textbf{10}, 270-276 (2015).
\bibitem{13}J. Jeong, N. Aetukuri, T. Graf, T. D. Schladt, M. G. Samant, and S. S. P. Parkin, Science
\textbf{339}, 1402-1405 (2013).
\bibitem{14}T. D. Schladt, T. Graf, N. B. Aetukuri, M. Y. Li, A. Fantini, X. Jiang, M. G. Samant, and S. S. P. Parkin, ACS Nano
\textbf{7}, 8074-8081 (2013).
\bibitem{15}B. Lei, N. Z. Wang, C. Shang, F. B. Meng, L. K. Ma, X. G. Luo, T. Wu, Z. Sun, Y. B. Zhang, and X. H. Chen, arXiv:1609.07726.
\bibitem{16}A. Tamai, A. Y. Ganin, E. Rozbicki, J. Bacsa, W. Meevasana, P. D. C. King, M. Caffio, R. Schaub, S. Margadonna, K. Prassides, M. J. Rosseinsky, and F. Baumberger, Phys. Rev. Lett.
\textbf{104}, 097002 (2010).
\bibitem{17}E. Ieki, K. Nakayama, Y. Miyata, T. Sato, H. Miao, N. Xu, X.-P. Wang, P. Zhang, T. Qian, P. Richard,Z. J. Xu, J. S. Wen, G. D. Gu, H. Q. Luo, H. H. Wen, H. Ding, and T. Takahashi, Phys. Rev. B
\textbf{89}, 140506(R) (2014).
\bibitem{18}R. Khasanov, M. Bendele, A. Amato, P. Babkevich, A. T. Boothroyd, A. Cervellino, K. Conder, S. N. Gvasaliya, H. Keller, H.-H. Klauss, H. Luetkens, V. Pomjakushin, E. Pomjakushina, and B. Roessli, Phys. Rev. B
\textbf{80}, 140511(R) (2009).
\bibitem{19}M. Bendele, S. Weyeneth, R. Puzniak, A. Maisuradze, E. Pomjakushina, K. Conder, V. Pomjakushin, H. Luetkens, S. Katrych, A. Wisniewski, R. Khasanov, and H. Keller, Phys. Rev. B
\textbf{81}, 224520 (2010).
\bibitem{20}V. Tsurkan, J. Deisenhofer, A. Gunther, Ch. Kant, M. Klemm, H.-A. Krug von Nidda, F. Schrettle, and A. Loidl, Eur. Phys. J. B
\textbf{79}, 289-299 (2011).
\bibitem{21}D. J. Gawryluk, J. Fink-Finowicki , A. Wisniewski, R. Puzniak , V. Domukhovski, R Diduszko, M Kozowski and M Berkowski, Supercond. Sci. Technol.
\textbf{24}, 065011 (2011).
\bibitem{22}C. H. Wu, W. C. Chang, J. T. Jeng, M. J. Wang, Y. S. Li, H. H. Chang, and M. K. Wu, Appl. Phys. Lett.
\textbf{102}, 222602 (2013).
\bibitem{23}M. Bendele, Z. Guguchia, F. von Rohr, T. Irifune, T. Shinmei, I. Kantor, S. Pascarelli, B. Joseph, and C. Marini, Phys. Rev. B
\textbf{90}, 174505 (2014).
\bibitem{24}B. Lei, J. H. Cui, Z. J. Xiang, C. Shang, N. Z. Wang, G. J. Ye, X. G. Luo, T. Wu, Z. Sun, and X. H. Chen, Phys. Rev. Lett.
\textbf{116}, 077002 (2016).
\bibitem{25}J. G. Guo, S. F. Jin, G. Wang, S. C. Wang, K. X. Zhu, T. T. Zhou, M. He, and X. L. Chen, Phys. Rev. B
\textbf{82}, 180520(R) (2010).
\bibitem{26}Y. Sun, T. Yamada, S. Pyon, and T. Tamegai, Sic. Rep.
\textbf{6}, 32290 (2016).
\bibitem{27}Y. Sun, Y. Tsuchiya, T. Taen, T. Yamada, S. Pyon, A. Sugimoto, T. Ekino, Z. X. Shi, and T. Tamegai, Sci. Rep.
\textbf{4}, 4585 (2014).
\bibitem{28}P. H. Lin, Y. Texier, A. Taleb-Ibrahimi, P. LeFevre, F. Bertran, E. Giannini, M. Grioni, and V. Brouet, Phys. Rev. Lett.
\textbf{111}, 217002 (2013).



\end{thebibliography}
\end{document}